\documentclass[twocolumn,preprintnumbers,,superscriptaddress,amsmath,amssymb]{revtex4}
\usepackage{amsfonts}
\usepackage{epsf}
\usepackage{anysize}
\marginsize{0.8 in}{0.8 in}{1in}{1.4in}
\usepackage{float}
\usepackage{graphicx}
\usepackage{dcolumn}
\usepackage{bm}
\usepackage{amsmath}
\usepackage{amssymb}
\usepackage{mathrsfs}
\usepackage{pstricks,pst-node,pst-text,pst-3d}
\usepackage{subfigure}
\usepackage[all]{xy}
\usepackage{color}
\usepackage{epsf}
\usepackage{anysize}
\usepackage{graphicx}
\usepackage{dcolumn}
\usepackage{bm}
\usepackage{amsmath}
\usepackage{amssymb}
\usepackage{mathrsfs}
\usepackage{pstricks,pst-node,pst-text,pst-3d}
\usepackage[all]{xy}
\usepackage{mathtools}

\DeclarePairedDelimiter\ket{\lvert}{\rangle}
\DeclarePairedDelimiterX\braket[2]{\langle}{\rangle}{#1 \delimsize\vert #2}

\usepackage[unicode=true,pdfusetitle,
 bookmarks=true,bookmarksnumbered=false,bookmarksopen=false,
 breaklinks=false,pdfborder={0 0 0},pdfborderstyle={},backref=false,colorlinks=true]
 {hyperref}
\hypersetup{
 colorlinks,linkcolor=red,citecolor=blue}

\makeatletter
  \newcommand\figcaption{\def\@captype{figure}\caption}
  \newcommand\tabcaption{\def\@captype{table}\caption}
\makeatother

{\tiny }

\begin{document}


\title{Flying-Qubit Control via a Three-level Atom with Tunable Waveguide Couplings}
\author{Wenlong Li}
\affiliation{Center for Intelligent and Networked Systems, Department of Automation, Tsinghua University, Beijing 100084, China}

\author{Xue Dong}
\affiliation{Center for Intelligent and Networked Systems, Department of Automation, Tsinghua University, Beijing 100084, China}

\author{Guofeng Zhang}
\thanks{E-mail: guofeng.zhang@polyu.edu.hk}
\affiliation{Department of Applied Mathematics, The Hong Kong Polytechnic University, Hong Kong, China}

\author{Re-Bing Wu}
\thanks{E-mail: rbwu@tsinghua.edu.cn}
\affiliation{Center for Intelligent and Networked Systems, Department of Automation, Tsinghua University, Beijing 100084, China}


\begin{abstract}
The control of flying qubits is at the core of quantum networks. As often carried by single-photon fields, the flying-qubit control involves not only their logical states but also their shapes. In this paper, we explore a variety of flying-qubit control problems using a three-level atom with time-varying tunable couplings to two input-output channels. It is shown that one can tune the couplings of a $\Lambda$-type atom to distribute a single photon into the two channels with arbitrary shapes, or use a $V$-type atom to catch an arbitrary-shape distributed single photon. The $\Lambda$-type atom can also be designed to transfer a flying qubit from one channel to the other, with both the central frequency and the photon shape being converted. With a $\Xi$-type atom, one can use the tunable coupling to shape a pair of correlated photons via cascaded emission. In all cases, analytical formulas are derived for the coupling functions to fulfil these control tasks, and their physical limitations are discussed as well. {These results provide useful control protocols for high-fidelity quantum information transmission over complex quantum networks.}
\end{abstract}

\maketitle

%
\section{Introduction}
The engineering of flying qubits is fundamentally important in coherent information transmission over quantum networks~\cite{DiVincenzo1995,Cirac1997,Kimble2008,Kuhn2002,Duan2001,Gheri1998}. Usually, the logical states of flying qubits are encoded by the number or polarization of photons contained in a pulsed electromagnetic field. Moreover, since the photon field often involves a continuous band of modes, its spectrum (or equivalently its temporal shape) is also to be controlled so as to physically match the receiver quantum nodes.

The control of flying qubits must be actuated by some manipulatable standing quantum system (i.e., an atom). From the viewpoint of quantum input-output theory~\cite{Gardiner1985,Combes2017}, the incoming and outgoing flying qubits form the quantum input and output of the standing quantum system, leading to the following three classes of control tasks: (1) the generation of flying qubits with vacuum quantum inputs~\cite{Kurtsiefer2002,Houck2007, Pierre2014,Gough2015,Trivedi2018,Fischer2018}, (2) the catching of flying qubits with vacuum quantum outputs~\cite{Stobinska2009,Wang2011,Yin2013,Aljunid2013,Pierre2014,Nurdin2016,Sounas2020}, and (3) the transformation of flying qubits with quantum inputs and outputs both non-vacuum ~\cite{Zhang2014,Zhang2017,Leong2016,Trivedi2018,Hurst2018,Kiilerich2019}.

Depending on the purposes of flying-qubit control, the actuating standing quantum system may have various structures. For the generation or catching of arbitrary-shape single flying qubits, it is sufficient to apply a simple two-level atom~\cite{Nurdin2016,Li2020,Li2021}. However, when processing multiple flying qubits or converting flying qubits between different channels, multi-level atoms are required~\cite{Pan2017,Sch_ll_2020,Sbresny2022}. As the simplest extension, a three-level atom can be simultaneously coupled to two channels, which enables the conversion of flying qubits from one channel to the other or the generation of an entangled pair of flying qubits.

In superconducting quantum circuits, all kinds of three-level artificial atoms can be realized with tunable frequency and coupling strength~\cite{Liu2005,Peng2016,Gu2017}, including the $\Lambda-$type~\cite{Vladimir2009}, the $\Xi-$type~\cite{Abdumalikov2010}, the $V-$type~\cite{Srinivasan2011} and the cyclic $\Delta-$type~\cite{Liu2005} that does not exist in natural atoms. However, little attention has been paid to the flying-qubit control with such systems. To our knowledge, $\Lambda$-type atoms were proposed for the generation of persistent single photons~\cite{You2007,Astafiev2007}, the nonreciprocal routing of single photons among two coupled chiral waveguides~\cite{Gonzalez2016} and the few-photon scattering processes~\cite{Hurst2018}, while $\Xi$-type atoms can be taken as transistors for switching single photons scattered by the atom~\cite{Kyriienko2016}. One can also apply the $\Delta$-type atom to the generation of microwave photon pairs~\cite{Marquardt2007,Peng2015} and the routing of single photons~\cite{Zhou2013}.

There was also not much attention paid to the shape control of continuous-mode flying qubits~\cite{Forn-Diaz2017}. Since the underlying control processes often involve time-varying parameters, the broadly used frequency-domain scattering analysis are not applicable as it can only deal with static systems. In the literature, various approaches have been proposed to model such dynamical processes, including the master equations~\cite{Scully1997}, quantum Langevin equations~\cite{Gardiner1985}, quantum trajectories~\cite{Carmichael1993,Gough2014}, quantum scattering theory~\cite{Zhang2014,Shi2015,Zhang2017,Trivedi2018,Fischer2018}, or quantum stochastic differential equations (QSDEs)~\cite{Li2020}.

In this paper, we will adopt the QSDE-based approach to explore a variety of flying-qubit control problems with a three-level atom~\cite{Liu2014} whose couplings to two quantum input-output channels are tunable. Analytic conditions will be derived for the coupling functions, as well as the time-varying detuning functions, to generate, catch or convert flying qubits with different types of three-level atoms, and their physical limitations will be discussed as well.

The remainder of this paper is organized as follows. Section~\ref{Sec:Model} describes the flying-qubit control model based on our previous work~\cite{Li2020,Li2021}. In Sec.~\ref{Sec:Generation}, we design control protocols for the generation of entangled flying qubits and cascaded generation of flying qubit pairs, following which the catching and conversion of flying qubits are discussed in Sec.~\ref{Sec:Catching} and Sec.~\ref{Sec:Transformation}, respectively. Finally, conclusion is drawn in Sec.~\ref{Sec:Conclusion}.

\section{The Model of flying-qubit control systems}\label{Sec:Model}
In this section, we will introduce the representation of flying-qubit states and their joint state with the standing three-level quantum system. Then, we will provide the formulation for calculating outgoing flying-qubit states under time-dependent controls.

\begin{figure}	\centering
	\includegraphics[width=0.8\columnwidth]{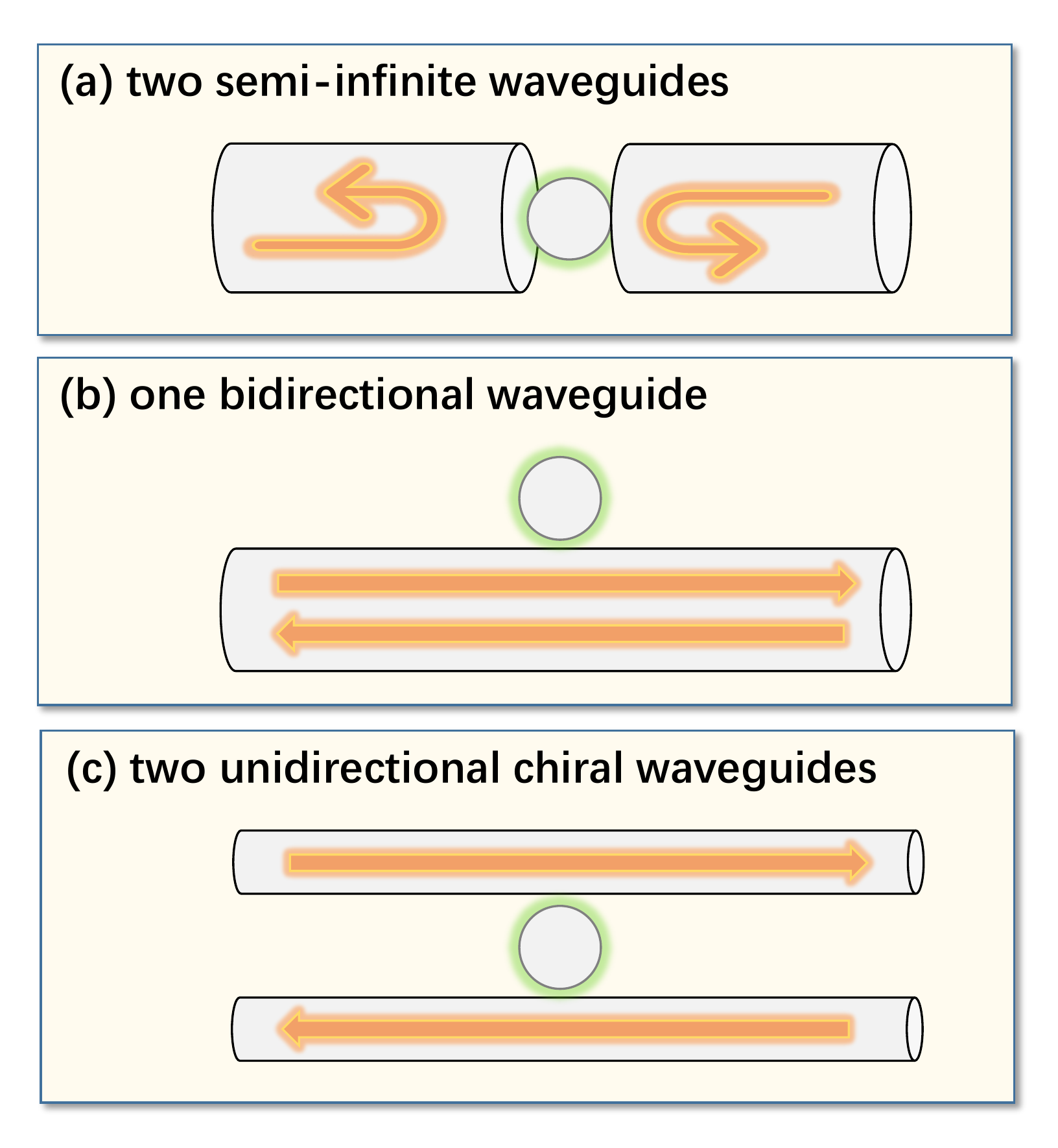}
	\caption{Schematics of a standing quantum system that is coupled to two quantum input-output channels, where the channels may physically correspond to (a) two semi-infinite waveguides, (b) one bidirectional waveguide, or (c) two unidirectional chiral waveguides.} \label{fig:1}
\end{figure}

\subsection{The representation of flying-qubit states}
Flying qubits are usually carried by traveling photon fields, and their logical states are encoded by the number of photons contained in the field, i.e., the logical states $\ket{0}$ and $\ket{1}$ are represented by the field's vacuum state $\ket{\rm vac}$ and single-photon state $\ket{1_\xi}$. Here, the shape function $\xi(t)$ is an intrinsic property of the flying qubit in addition to the superposition of logical states, because only properly shaped photon fields can be perfectly received by remote quantum systems~\cite{Wang2011,Aljunid2013,Nurdin2016}.

To define the single-photon state $\ket{1_\xi}$ with shape function $\xi(t)$, let us start with the temporal annihilation operator $b(t)=\int_{-\infty}^{+\infty}e^{-i\omega t}b(\omega){\rm d}\omega$ of the traveling field in a waveguide, where $b(\omega)$ is the annihilation operator associated with mode $\omega$. The state of a flying qubit that contains exactly one photon in a single channel can be described as~\cite{Zhang2017,Zhang2021}
\begin{equation}\label{eq:a single-photon state in one channel}
|1_{\xi}\rangle =\int_{-\infty}^{+\infty} \xi(\tau) b^{\dagger}(\tau)|{\rm vac}\rangle {\rm d}\tau,
\end{equation}
where the shape function $\xi(t)$ represents the probability density amplitude of finding a photon at moment $t\in(-\infty,\infty)$, and the preservation of the overall probability requires that $\xi(t)$ be normalized, i.e., $\int_{-\infty}^{+\infty}|\xi(\tau)|^2{\rm d}\tau=1$.

The single-photon field can also be distributed in multiple channels, which can be represented by the following superposition of $m$ single-photon components:
\begin{equation}\label{eq:a single-photon state in two channel}
|1_{\xi}\rangle =\sum_{j=1}^m \int_{-\infty}^{+\infty}  \xi_j(\tau) b^{\dagger}_j(\tau)|{\rm vac}\rangle {\rm d}\tau,
\end{equation}
where $b_j^{\dagger}(\tau)$ and $\xi_j(\tau)$ are the field operator and photon shape function in the $j$th channel, respectively.

More generally, we can define $n$-photon states in $m$ channels as follows
\begin{widetext}
	\begin{equation}\label{eq:n photons state in two channel}
	|n_{\xi}\rangle =\sum_{j_1,\cdots,j_n=1}^m \int_{-\infty}^{+\infty} {\rm d}\tau_n \int_{-\infty}^{\tau_n} {\rm d}\tau_{n-1} \cdots   \int_{-\infty}^{\tau_2} {\rm d}\tau_1 \ \xi_{j_1,\cdots,j_n}(\tau_1,\cdots,\tau_n)
	b^{\dagger}_{j_n}(\tau_n) \cdots b^{\dagger}_{j_1}(\tau_1) |{\rm vac}\rangle,
	\end{equation}
\end{widetext}
where the shape function $ \xi_{j_1,\cdots,j_n}(\tau_1,\cdots,\tau_n)$ indicates the probability density amplitude of generating the $k$th photon in the $j_k$th channel at time $\tau_k$. For notational consistence, we denote the shape function with zero photon as $\xi$ that does not have any superscripts or subscripts.

\subsection{The computation of outgoing flying-qubit states}

Consider a three-level atom that is coupled to two quantum input-output channels. Let $|g\rangle$, $|e\rangle$ and $|f\rangle$ be the atom's eigenstates, where $|f\rangle$ is the highest excited state, and $\mathcal{H}$ is the Hilbert space they span. As is shown in Fig.~\ref{fig:1}, the two channels can be physically realized by two semi-infinite waveguides, a bidirectional waveguide or two unidirectional waveguides. Throughout this paper, we take the third scenario for the description of problem setup, but the obtained results can be directly extended to the other two scenarios.

To analyze the underlying quantum input-output processes, we generalize the above flying-qubit state representation to the joint state of the interacting atom-waveguide system, as follows:
\begin{widetext}
	\begin{equation}\label{eq:the joint system-field state}
	\begin{split}
	|\Psi(t)\rangle&=\sum_{n=0}^{\infty} \quad \sum_{j_1,\cdots,j_n=1}^2  \int_{-\infty}^t {\rm d}\tau_n \int_{-\infty}^{\tau_n} {\rm d}\tau_{n-1}\cdots   \int_{-\infty}^{\tau_2}{\rm d}\tau_1 \ |\psi_{j_1,\cdots,j_n}(\tau_1,\cdots,\tau_n|t)\rangle\otimes
	b^{\dagger}_{j_n}(\tau_n) \cdots b^{\dagger}_{j_1}(\tau_1) |{\rm vac}\rangle,
	\end{split}
	\end{equation}
\end{widetext}
where $|\psi_{j_1,\cdots,j_n}(\tau_1,\cdots,\tau_n|t)\rangle\in\mathcal{H}$ is the atom's associated state at time $t$ when $n$ photons are observed at moments $\tau_1\leq\cdots\leq\tau_n$ in the $j_1$th, $\cdots$, $j_n$th channels, respectively. Note that each $|\psi_{j_1,\cdots,j_n}(\tau_1,\cdots,\tau_n|t)\rangle$ is generally unnormalized, but the total state $|\Psi(t)\rangle$ is always normalized.

Assume that the quantum input field (i.e., incoming flying qubits) is in the vacuum state $|{\rm vac}\rangle$, and the atom is initially prepared at state $|\psi(-\infty)\rangle=|\psi_0\rangle$. In Refs.~\cite{Li2020,Li2021}, it is proven that the time evolution of the joint state $\ket{\Psi(t)}$ is determined by a non-unitary evolution operator $V(t)$ on $\mathcal{H}$, which is governed by the following equation~\cite{Gardiner2004}
\begin{equation}\label{eq:the effictive H}
\dot{V}(t)=\left[-iH(t) -\frac{1}{2}\sum_{j=1}^2 L^{\dagger}_j(t)L_j(t)\right]V(t),
\end{equation}
where $V(-\infty)=\mathbb{I}$, $H(t)$ is the atom's Hamiltonian including its internal part and its interaction part with external coherent control fields, and $L_i(t)$ ($i=1,2$) is its coupling operator to the $j$th channel. Our previous studies showed that each $|\psi_{j_1,\cdots,j_n}(\tau_1,\cdots,\tau_n|t)\rangle$ can be calculated as follows:
\begin{equation}\label{eq:the solution of the equivalent state of system}
\begin{split}
  &\quad|\psi_{j_1,\cdots,j_n}(\tau_1,\cdots,\tau_n|t)\rangle
\\&=V(t)\widetilde{L}_{j_n}(\tau_n)\cdots\widetilde{L}_{j_1}(\tau_1) |\psi_0)\rangle,
\end{split}
\end{equation} 	
where $\widetilde{L}_{j_i}(\tau_i)=V^{-1}(\tau_i)L_{j_i}(\tau_i)V(\tau_i)$.

The outgoing flying-qubit state can be extracted from the asymptotic limit of $|\Psi(t)\rangle$ when $t\rightarrow\infty$. Owing to the decaying nature of $V(t)$, each component  $\quad|\psi_{j_1,\cdots,j_n}(\tau_1,\cdots,\tau_n|t)\rangle$ must decay to either $|g\rangle$ or $|e\rangle$, and can thus be decomposed as:
\begin{align}\label{eq:the final state}
&|\psi_{j_1,\cdots,j_n}(\tau_1,\cdots,\tau_n|\infty)\rangle \nonumber\\
=&~\xi^g_{j_1,\cdots,j_n}(\tau_1,\cdots,\tau_n)|g\rangle+\xi^e_{j_1,\cdots,j_n}(\tau_1,\cdots,\tau_n)|e\rangle.
\end{align}
Taking the asymptotic limit, we can reexpress the joint state as
\begin{widetext}
	\begin{eqnarray}\label{eq:the joint system-field state at infty}
	|\Psi(\infty)\rangle&=&|g\rangle \otimes\left[\sum_{n=0}^{\infty}  \sum_{j_1,\cdots,j_n}\int_{-\infty}^t {\rm d}\tau_n \cdots   \int_{-\infty}^{\tau_2}{\rm d}\tau_1
	\xi^g_{j_1,\cdots,j_n}(\tau_1,\cdots,\tau_n) b^{\dagger}_{j_n}(\tau_n) \cdots b^{\dagger}_{j_1}(\tau_1) |{\rm vac}\rangle\right]\nonumber \\
&&+|e\rangle \otimes\left[\sum_{n=0}^{\infty}  \sum_{j_1,\cdots,j_n}\int_{-\infty}^t {\rm d}\tau_n \cdots   \int_{-\infty}^{\tau_2}{\rm d}\tau_1
	\xi^e_{j_1,\cdots,j_n}(\tau_1,\cdots,\tau_n) b^{\dagger}_{j_n}(\tau_n) \cdots b^{\dagger}_{j_1}(\tau_1) |{\rm vac}\rangle\right],
	\end{eqnarray}
\end{widetext}
in which the summations in the brackets represent the state of the emitted photon field when the atom decays $|g\rangle$ or $|e\rangle$. When the number of excitation numbers is preserved, the summations usually contain a finite number of terms and hence can be effectively calculated to analyze the control processes.

\begin{figure}
	\centering
	\includegraphics[width=1\columnwidth]{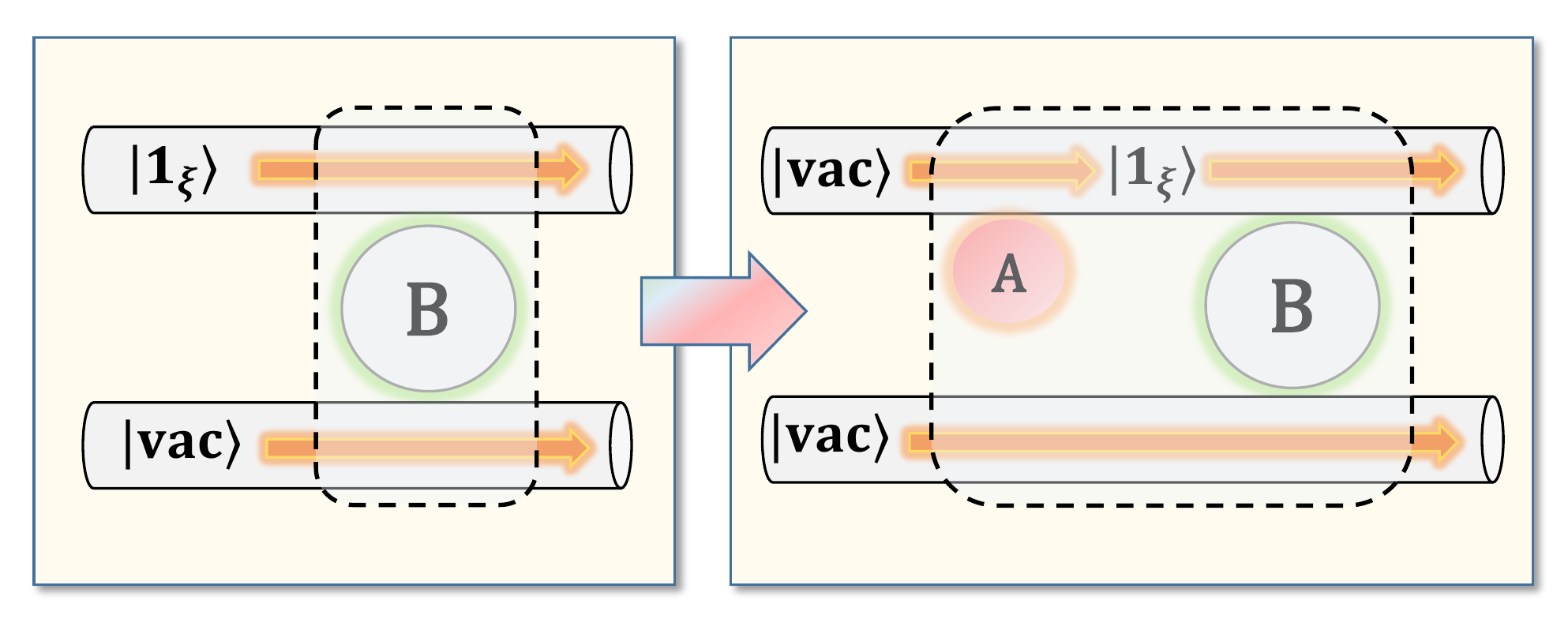}
	\caption{Schematics for transforming a non-vacuum input system (the box on the left) to an equivalent system with vacuum input (the box on the right). The single-photon input state $\ket{1_\xi}$ is generated by a properly designed ancillary system $A$.} \label{fig:2}
\end{figure}

When the quantum input of some channel is non-empty (i.e., the field is initially at a non-vacuum state), we can generalize the above procedure by cascading an ancillary system that generates the quantum input. As is shown in Fig.~\ref{fig:2}, the cascaded system receives vacuum quantum inputs, to which the above procedure can be directly applied.

Suppose that the non-vacuum quantum input is fed into the first channel. Let $H_A(t)$ and $L_A(t)$ be the Hamiltonian and the coupling operator of the ancillary system $A$, which can be artificially chosen according to the desired input state $|1_\xi\rangle$. According to the $(S,L,H)$ formula~\cite{Gough2009}, the equivalent Hamiltonian of the joint system is
\begin{eqnarray}\label{eq:the equivalent Hamiltonian of the joint system}
\bar{H}(t)&=&H_A(t)\otimes \mathbb{I}+\mathbb{I}_A\otimes H(t)\nonumber \\
&&+\frac{1}{2i}\left[L_{1}^{\dagger}(t)L_{A}(t)-L_{A}^{\dagger}(t)L_{1}(t)\right].
\end{eqnarray}
and the equivalent coupling operators are
\begin{subequations}\label{eq:the equivalent coupling operators of the joint system}
 \begin{align}
 \bar{L}_1(t) =& ~\mathbb{I}_A\otimes L_{1}(t) +L_{A}(t)\otimes \mathbb{I}, \\
 \bar{L}_{2}(t) =&~ \mathbb{I}_A\otimes L_2(t).
		\end{align}
\end{subequations}
The equivalent $\bar{H}(t)$ and $\bar{L}_j(t)$ can then be applied to construct the nonunitary evolution equation (\ref{eq:the effictive H}) for the calculation of output fields. The same operations can be done when the second channel or both channels have non-vacuum inputs.

\subsection{Existing results on flying-qubit control with two-level atoms}
Before studying flying-qubit control actuated by a three-level atom, we briefly review the known results with a two-level atom.

Suppose that the internal Hamiltonian of the two-level system is $H(t)=\epsilon(t)\sigma_z$, where $\epsilon(t)$ is the detuning between the atomic transition frequency and the central frequency of the field in the channel. The coupling operator is $L(t)=\sqrt{\gamma(t)}\sigma_-$, where $\sigma_-$ is the standard Pauli lowering operator. In previous works~\cite{Li2020,Li2021}, it is derived that, to generate a single-photon pulse $\xi(t)=|\xi(t)|e^{-i\phi(t)}$ from an excited two-level atom,  the coupling function should be set to
\begin{equation}\label{eq:the conditions about gamma when generating by 2-level}
\bar{\bar{\gamma}}(t)=\frac{|\xi(t)|^2}{\int_t^{\infty}|\xi(s)|^2{\rm d}s},
\end{equation}
and the detuning frequency needs to match the phase condition $\epsilon(t)=\dot{\phi}(t)$, where $\phi(t)$ must start from $\phi(-\infty)=0$.  Hereafter, the double bars over $\gamma(t)$ are used to denote coupling functions associated with two-level atoms.

One can also derive that, to catch a single photon $|1_\xi\rangle$ with $\xi(t)=|\xi(t)|e^{-i\phi(t)}$, the coupling function $\gamma(t)$ should be set to
\begin{equation}\label{eq:the conditions about gamma when catching by 2-level}
\bar{\bar{\gamma}}(t)=\frac{|\xi(t)|^2}{\int_{-\infty}^{t}|\xi(s)|^2{\rm d}s},
\end{equation}
while the same phase condition needs to be matched.

The above conditions for controlling flying qubits exhibit an elegant symmetry in that the integral in the denominator points to the future when generating a single photon and to the past when catching a single photon. Owing the preservation of energy, it can be verified that these two denominators both equals the population of the excited state at the present time $t$. Later we will see that this rule also holds for control protocols using a three-level atom.

\section{The Generation of Flying Qubits with a Three-level Atom}\label{Sec:Generation}
In this section, we will use the above flying-qubit control model to design control functions for generating flying qubits with a $\Lambda$-type or a $\Xi$-type atom.

 \subsection{Entangled flying qubits generated by a $\Lambda$-type atom}\label{Sec:GenerationA}
 \begin{figure}
 	\centering
 	\includegraphics[width=0.9\columnwidth]{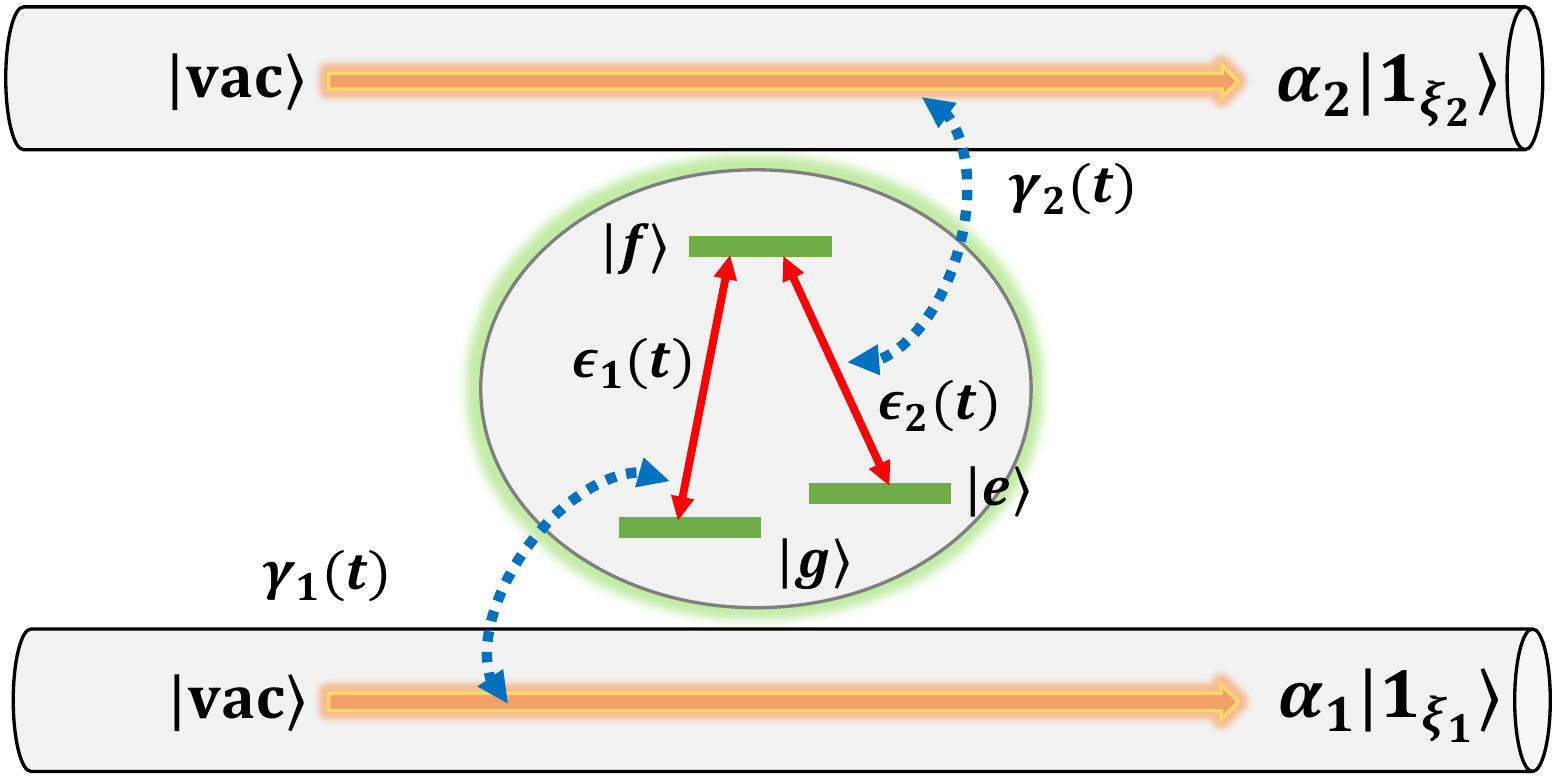}
 	\caption{Schematics for the generation of a distributed single photon with a $\Lambda$-type three-level atom, which forms a pair of flying qubits that is entangled with the atom at state $\alpha_1|g\rangle|1_{\xi_1}\rangle|{\rm vac}\rangle +\alpha_2|e\rangle|{\rm vac}\rangle|1_{\xi_2}\rangle $. Here, $\epsilon_1(t)$ and $\epsilon_2(t)$ are the detunings between the central frequency of the incident field and the corresponding atomic transition frequencies, respectively, and $\gamma_{1}(t)$ and $ \gamma_{2}(t)$ are the coupling functions.} \label{fig:3}
 \end{figure}

Consider a $\Lambda$-type atom that is coupled to two quantum channels (see Fig.~\ref{fig:3}).
Being initially excited to the state $\ket{f}$, the atom will emit a single photon distributed into the two channels, forming a pair of flying qubits that are entangled with the atom after they leave. Their joint state can be written as
\begin{equation}\label{eq:entangled state}
|\Psi(\infty)\rangle =\alpha_1|g\rangle\otimes|1_{\xi_1}\rangle|{\rm vac}\rangle +\alpha_2|e\rangle\otimes|{\rm vac}\rangle|1_{\xi_2}\rangle,
\end{equation}
where the coefficients $\alpha_{1}$ and $\alpha_2$ are complex numbers and the shape functions $\xi_{1}(t)$ and $\xi_2(t)$ are normalized.

To generate such distributed single photons, we can apply the following coupling operators:
\begin{equation}
L_1(t)=\sqrt{\gamma_1(t)}|g\rangle\langle f|,\quad L_2(t)=\sqrt{\gamma_2(t)}|e\rangle\langle f|,
\end{equation}
where the tunable coupling functions $\gamma_1(t)$ and $\gamma_2(t)$ alter the instantaneous rate of field emission into the corresponding channels. In the rotating-wave frame, the system's Hamiltonian reads
\begin{equation}
H(t)=\left[\epsilon_1(t)+\epsilon_2(t)\right]|f\rangle\langle f|,
\end{equation}
where $\epsilon_1(t)$ and $\epsilon_2(t)$ are the detunings between the transition frequencies and central frequencies of the channels.

\begin{figure}
  	\centering
  	\includegraphics[width=1\columnwidth]{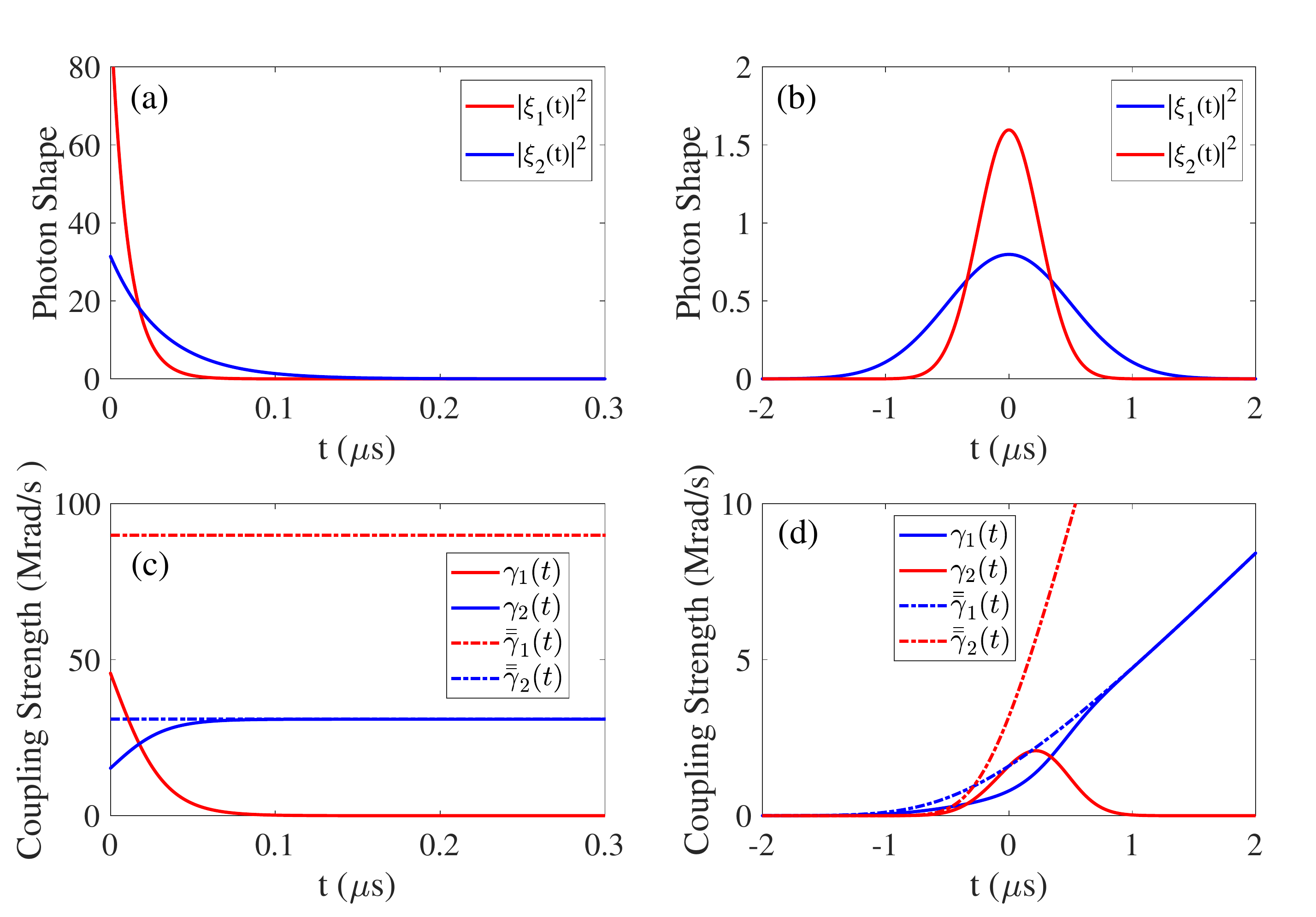}
  	\caption{The generation of a pair of flying qubits at the joint state as $\frac{1}{\sqrt{2}}(|g\rangle|1_{\xi_1}\rangle|{\rm vac}\rangle +|e\rangle|{\rm vac}\rangle|1_{\xi_2}\rangle )$  using a $\Lambda$-type atom. The output single photons have (a) exponentially decaying shapes or (b) Gaussian shapes, whose corresponding coupling strengths are shown in (c) and (d), respectively.} \label{fig:4}
  \end{figure}
According to Eqs.~(\ref{eq:the solution of the equivalent state of system})-(\ref{eq:the joint system-field state at infty}), we can derive that (see Appendix~\ref{app:generationA} for details), for arbitrary given coefficients $\alpha_1$ and $\alpha_2$ and shape functions $\xi_1(t)$ and $\xi_2(t)$, one can generate the entangled state (\ref{eq:entangled state}) using the following coupling functions
\begin{subequations}\label{eq:the conditions about gamma when generating by lambda}
\begin{align}
&\gamma_1(t)=\frac{|\alpha_1\xi_1(t)|^2}{\int_t^{\infty}|\alpha_1\xi_1(s)|^2{\rm d}s+\int_t^{\infty}|\alpha_2\xi_2(s)|^2{\rm d}s},
\\&\gamma_2(t)=\frac{|\alpha_2\xi_2(t)|^2}{\int_t^{\infty}|\alpha_1\xi_1(s)|^2{\rm d}s+\int_t^{\infty}|\alpha_2\xi_2(s)|^2{\rm d}s},
\end{align}
\end{subequations}
which have a common denominator because the coupling operators of the two channels share the excited state $|f\rangle$.

Let $\xi_j(t)=|\xi_j(t)|e^{-i\phi_j(t)}$ ($j=1,2$) with $\phi_j(t)$ being the phase of the shape function. To match the phases, the detuning frequencies should satisfy
\begin{equation}\label{eq:the conditions about epsilon when generating by lambda}
\epsilon_1(t)+\epsilon_2(t)=\dot{\phi}_1(t)=\dot{\phi}_2(t),
\end{equation}
where $\phi_1(-\infty)=\phi_2(-\infty)=0$. This actually requires that $\xi_1(t)$ and $\xi_2(t)$ must have identical phase functions $\phi_1(t)=\phi_2(t)$.

The obtained solution (\ref{eq:the conditions about gamma when generating by lambda}) is closely related to the single-photon generation condition (\ref{eq:the conditions about gamma when generating by 2-level}) with a two-level atom. They are similar in that both denominators represent the total number of photons to be released in the future. Because the number of excitation numbers (or energy) is preserved, the denominators are also equal to the remaining population of the initially excited state ($|e\rangle$ for the two-level atom and $|f\rangle$ for the three-level atom) at present time $t$.

For illustration, we simulate two typical classes of single-photon shape functions. In the first case, we assume that both $\xi_1(t)$ and $\xi_2(t)$ are in the following exponentially decaying form
\begin{equation}\label{eq:the target exponential waveforms xi_1 and xi_2}
|\xi_k(t)|=\sqrt{\gamma_{ck}}e^{-\gamma_{ck} t/2},\quad k=1,2,
\end{equation}
where $\gamma_{c1}/2\pi =15$MHz and $\gamma_{c2} /2\pi =5$MHz. According to Eq.~(\ref{eq:the conditions about gamma when generating by lambda}), it is easy to calculate that, to generate a pair of arbitrary-shape flying qubits with the above waveforms at the joint state as $\frac{1}{\sqrt{2}}(|g\rangle|1_{\xi_1}\rangle|{\rm vac}\rangle +|e\rangle|{\rm vac}\rangle|1_{\xi_2}\rangle )$ , the coupling functions need to be
\begin{subequations}\label{eq:the conditions about gamma when generating by lambda when exponential waveforms}
\begin{align}
&\gamma_1(t)=\frac{\gamma_{c1}}{1+e^{(\gamma_{c1}-\gamma_{c2k})t}},
\\&\gamma_2(t)=\frac{\gamma_{c2}}{1+e^{(\gamma_{c2}-\gamma_{c1})t}}.
\end{align}
\end{subequations}

Recall that, according to Eq.~(\ref{eq:the conditions about gamma when generating by 2-level}), the coupling function for generating the two exponentially decaying waveforms with two-level atoms are ${\bar{\bar{\gamma}}}_1(t)\equiv\gamma_{c1}$ and ${\bar{\bar{\gamma}}}_2(t)\equiv\gamma_{c2}$, respectively. There is a competition between the photon emission in the two channels in that $\gamma_2(t)$ asymptotically approaches to ${\bar{\bar{\gamma}}}_2(t)$ but the more rapidly decaying $\gamma_1(t)$ gradually vanishes.

This competition also exists when the two shapes are both Gaussian functions:
\begin{equation}\label{eq:the target Gaussian pulse waveforms xi_1 and xi_2}
|\xi_k(t)|=\left(\frac{\Omega_k^2}{2\pi}\right)^{\frac{1}{4}}e^{-(\frac{\Omega_kt}{2})^2},\quad k=1,2,
\end{equation}
where $\Omega_{1}=2$MHz and $\Omega_2=4$MHz are the widths of the shapes. We can see in Fig~\ref{fig:4}(d) that, as predicted, the required coupling function $\gamma_1(t)$ approaches to ${\bar{\bar{\gamma}}}_1(t)$ because $\xi_1(t)$ decays more slowly, and $\gamma_2(t)$ gradually vanishes.

 \subsection{Cascaded generation of correlated flying qubits by a $\Xi$-type atom}\label{Sec:GenerationB}
  \begin{figure}
 	\centering
 	\includegraphics[width=0.8\columnwidth]{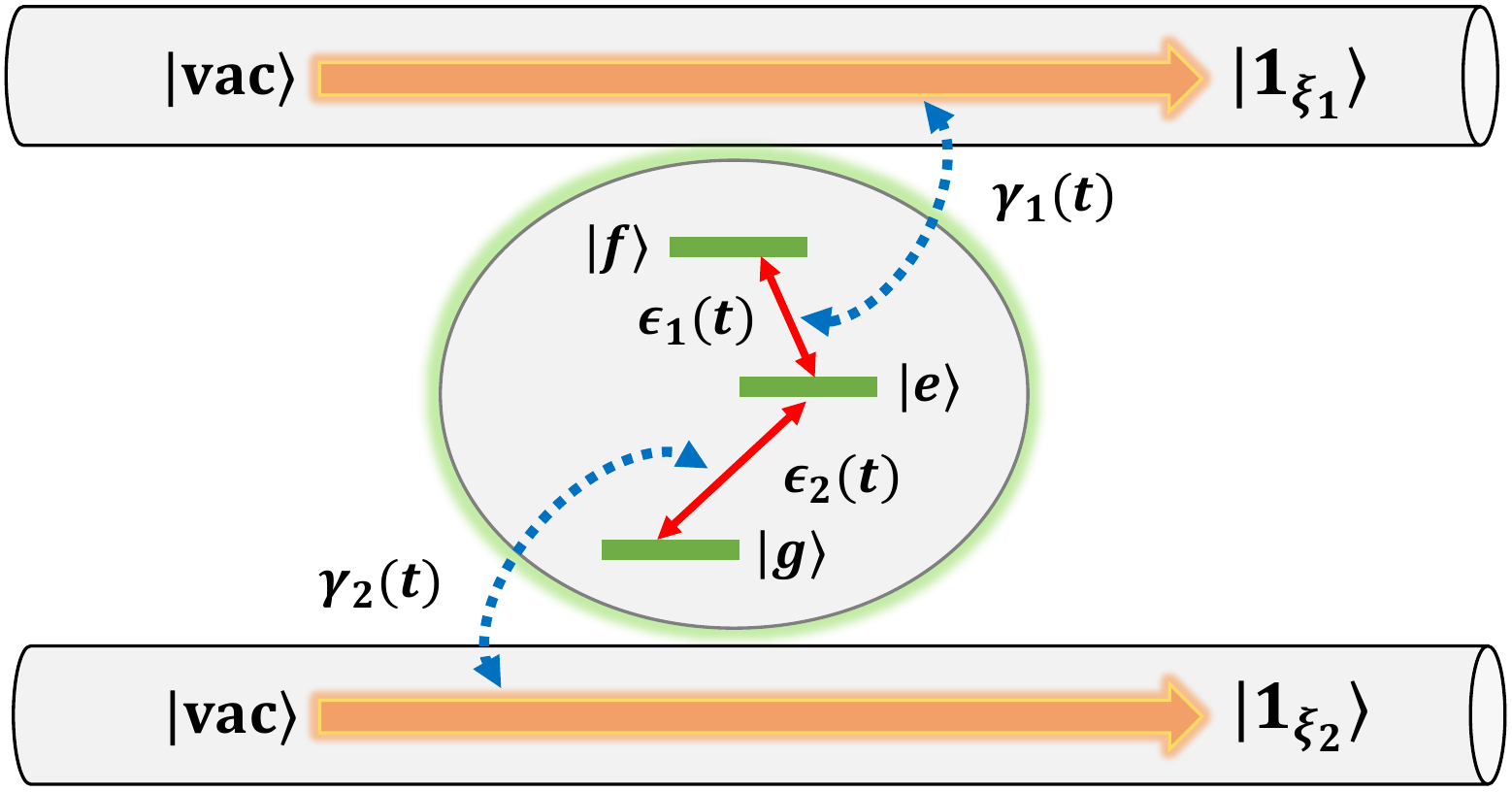}
 	\caption{Schematics for the generation of a pair of correlated single photons by a $\Xi$-type atom, including $|1_{\xi_1}\rangle$ in the 1st channel and $|1_{\xi_2}\rangle$ in the 2nd channel. Here, $\epsilon_1(t)$ and $\epsilon_2(t)$ are the detuning frequencies between the incident field and the $|f\rangle \leftrightarrow |e\rangle$ and $|e\rangle \leftrightarrow |g\rangle$ transistion frequencies, respectively. $\gamma_{1}(t)$ and $ \gamma_{2}(t)$ are the coupling coefficients between the $\Xi$-type atom and the channels.} \label{fig:5}
 \end{figure}

Consider a $\Xi$-type atom shown in Fig.~\ref{fig:5}. When the atom is initially prepared at the excited state $|f\rangle $, a pair of single photons will be sequentially emitted into the first and the second channels. Since the two photons are statistically correlated, their state cannot be decomposed as the direct product of two single-photon states like $|1_{\xi_1}\rangle\otimes |1_{\xi_2}\rangle$. Nevertheless, one can define the single-photon shape observed in each channel as the corresponding marginal probability distribution function:
\begin{eqnarray}\label{}
|\xi_1({\tau_1})|^2&=&\int_{\tau_1}^{\infty}|\xi_{1,2}^g(\tau_1,\tau_2)|^2{\rm d}\tau_2,\\
|\xi_2({\tau_2})|^2&=&\int_{-\infty}^{\tau_2}|\xi_{1,2}^g(\tau_1,\tau_2)|^2{\rm d}\tau_1,
\end{eqnarray}
where $\xi_{1,2}^g(t_1,t_2)$ is the shape of generated two-photon field that can be calculated from Eq.~(\ref{eq:the solution of the equivalent state of system}). In this case, $\xi_1(t)$ and $\xi_2(t)$ do not possess definite phases $\phi_1(t)$ and $\phi_2(t)$ as the two photons are not separable.

We seek proper coupling functions $\gamma_1(t)$ and $\gamma_2(t)$ to shape $|\xi_1(t)|^2$ and $|\xi_2(t)|^2$ as demanded. The underlying internal Hamiltonian and coupling operators of a $\Xi$-type atom are as follows:
\begin{eqnarray}
H(t)&=&\epsilon_1(t)|f\rangle\langle f|+\epsilon_2(t)|e\rangle\langle e|, \\
L_{1}(t)&=&\sqrt{\gamma_{1}(t)} |e\rangle\langle f|,\\
L_2(t)&=&\sqrt{\gamma_{2}(t)} |g\rangle\langle e|,
\end{eqnarray}
from which we can derive that (see Appendix~\ref{generator cascaded flying qubit} for details)
 \begin{subequations}\label{eq:the conditions about gamma when generating by Xi}
 	\begin{align}
 	& \gamma_1(t)=\frac{|\xi_1(t)|^2}{\int_{t}^{\infty}|\xi_1(s)|^2 {\rm d}s},
 	\\& \gamma_2(t)=\frac{|\xi_2(t)|^2}{\int_t^\infty|\xi_2(s)|^2{\rm d}s-\int_t^\infty|\xi_1(s)|^2 {\rm d}s}.
 	\end{align}
 \end{subequations}
The solution (\ref{eq:the conditions about gamma when generating by Xi}) indicates that $\gamma_1(t)$ is only dependent on the shape function $|\xi_1(t)|^2$ and thus its design can be treated as that with a two-level atom. However, the design of $\gamma_2(t)$ is not only dependent on the shape function $|\xi_2(t)|^2$ but also on $|\xi_1(t)|^2$ of the previously emitted photon.

The observed population rule in the above discussion can be verified here as the denominator of Eq.~(\ref{eq:the conditions about gamma when generating by Xi}a) is equal to the current population of $|f\rangle$, whose decay leads to the emission of the first photon. Also, owing to the conservation of energy, the denominator of Eq.~(\ref{eq:the conditions about gamma when generating by Xi}b) represents the accumulated population of the state $|e\rangle$ that is associated with the emission of the second photon, as it is equal to the difference between the energies dumped from the upper level $|f\rangle$ and to the lower level $g\rangle$ indicated by the denominator.

Due to the passivity of the control system, the energy dumped to $|g\rangle$ can only come from the energy dumped from $|f\rangle$, and hence the following physical realizability condition must be held:
\begin{equation}\label{eq:constraint}
  \int_t^\infty|\xi_2(s)|^2{\rm d}s>\int_t^\infty|\xi_1(s)|^2 {\rm d}s,\quad \forall t>0,
\end{equation}
which is also demanded by positivity of $\gamma_2(t)$ according to Eq.~ (\ref{eq:the conditions about gamma when generating by Xi}b). This condition implies that the tail area of $|\xi_2(t)|^2$ must be always greater than that of $|\xi_1(t)|^2$, and hence $|\xi_1(t)|^2$ should decay more quickly.

\begin{figure}
 	\centering
 	\includegraphics[width=1\columnwidth]{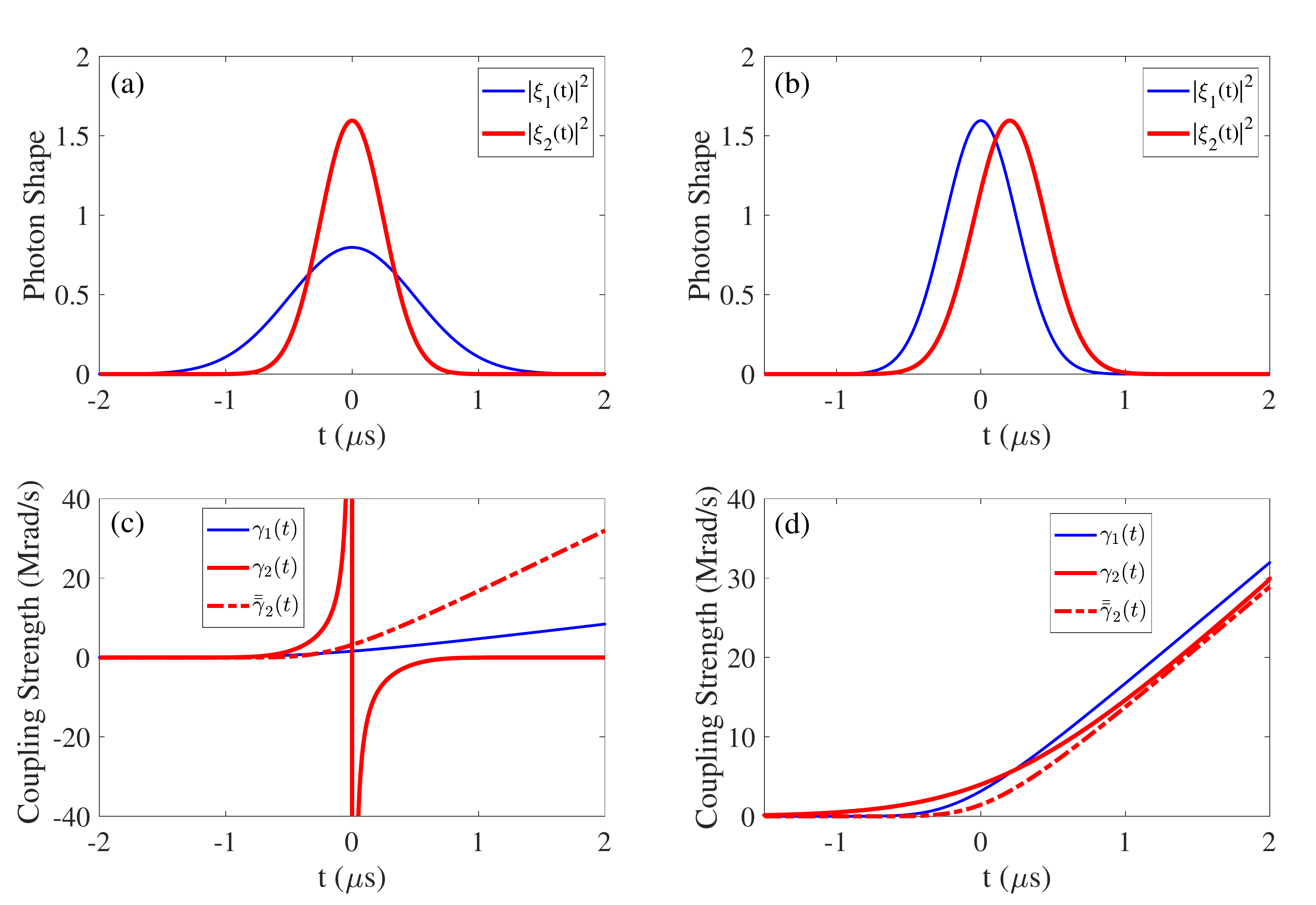}
 	\caption{The generation of a pair of correlated single photons $(|1_{\xi_1}\rangle|1_{\xi_2}\rangle$ using a $\Xi$-type atom. The incident photon shapes $\xi_1(t)$ and $\xi_2(t)$ are with (a) different Gaussian shapes or (b) same Gaussian shapes. The corresponding coupling strengths are shown in (c) and (d) respectively.} \label{fig:6}
 \end{figure}

We illustrate the above result with two representative examples of Gaussian photon pulses. In the first case, the two Gaussian pulses have different widths and identical peak times. As shown in Fig.~\ref{fig:6}(a) and (c), the tail areas of $|\xi_2(t)|^2$ is initially greater than that of $|\xi_1(t)|^2$, but it becomes smaller later. This leads to the singularity of $\gamma_2(t)$ at the peak time and non-physical negativity after that, which indicates that such photon shapes are not physically realizable by tunable couplings.

In the second example, the two photons have identical Gaussian shapes, but the second photon is behind the first with a time delay, as follows:
\begin{equation}\label{eq:the normal waveforms}
\begin{split}
&|\xi_1(t)|=\left(\frac{\Omega^2}{2\pi}\right)^{\frac{1}{4}}e^{-\left[\frac{\Omega t }{2}\right]^2},
\\&|\xi_2(t)|=\left(\frac{\Omega^2}{2\pi}\right)^{\frac{1}{4}}e^{-\left[\frac{\Omega(t-t_p)}{2}\right]^2},
\end{split}
\end{equation}
in which $t_p=0.2 {\rm \mu s}$. As are shown in Figs.~\ref{fig:6}(b) and (d),
$\gamma_2(t)$ is physically realizable because the condition (\ref{eq:constraint}) holds. Because the tail area of $|\xi_1(t)|^2$ decays more quickly, $\gamma_2(t)$ asymptotically approaches to ${\bar{\bar \gamma}}_2(t)$ that is associated to a two-level aotom.

\section{The catching of flying qubits by a three-level atom}\label{Sec:Catching}
\begin{figure}
 	\centering
 	\includegraphics[width=0.8\columnwidth]{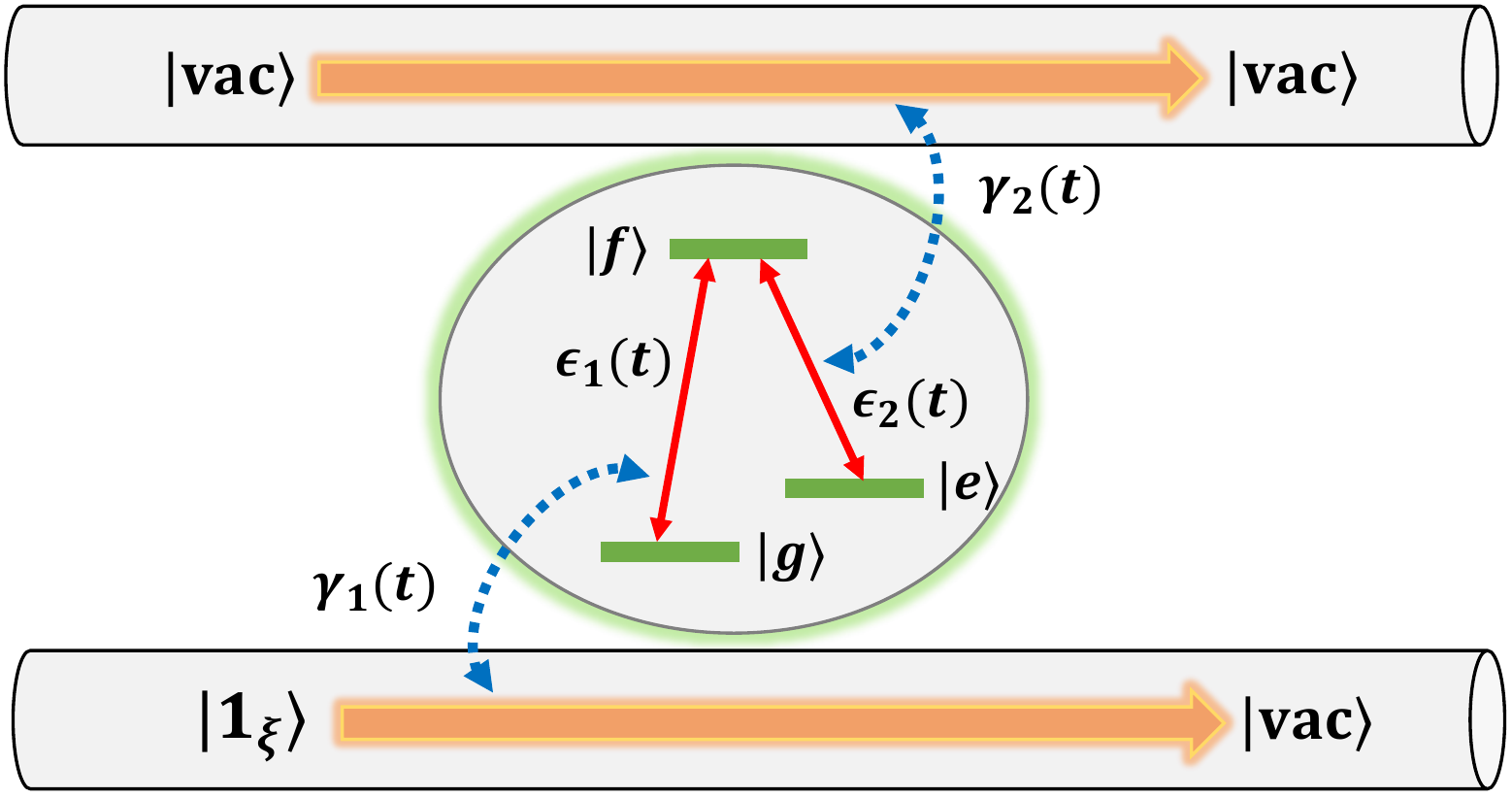}
 	\caption{Schematics for catching a flying qubit at state $|1_{\xi}\rangle|{\rm vac}\rangle$ by a $\Lambda$-type atom at state $|g\rangle$. Here, $\epsilon_1(t)$ and $\epsilon_2(t)$ are the detuning frequencies between the incident field and the $|f\rangle \leftrightarrow |g\rangle$ and $|f\rangle \leftrightarrow |e\rangle$ transistion frequencies, respectively. $\gamma_1(t)$ and $ \gamma_2(t)$ are the coupling coefficients between the $\Lambda$-type atom and the channels.} \label{fig:7}
 \end{figure}

Let us first consider the $\Lambda$-type atom shown in Fig.~\ref{fig:7}. When the atom is initially prepared at the state $|g\rangle $, and we wish to catch a single photon $|1_{\xi}\rangle$ from the first channel, i.e., finding proper control functions such that the ouputs of both channels are empty, and the atom's state transits from $|g\rangle$ to $|f\rangle$. Assuming that  $\xi(t)=|\xi(t)|e^{-i\phi(t)}$, we can introduce an ancillary system $A$ to generate $|1_\xi\rangle$, and apply Eqs.~(\ref{eq:the solution of the equivalent state of system})-(\ref{eq:the joint system-field state at infty}) to the joint standing system. It can be derived that the coupling function and detuning of the first channel must be
\begin{equation}\label{eq:the conditions about gamma when catch by lambda}
\gamma_1(t)=\frac{|\xi(t)|^2}{\int_{-\infty}^t|\xi(s)|^2 {\rm d}s}, \quad  	\epsilon_1(t)=\dot{\phi}(t).
\end{equation}	
Meanwhile, the coupling to the second channel must be turned off, i.e., $\gamma_2(t)\equiv 0$, so as to prevent the leakage of the photon into the second channel (The details of its derivation can see in Appendix~\ref{conversion cascaded flying qubit}).  Thus, the design of such controls is exactly the same as that with a two-level atom.

 \begin{figure}
 	\centering
 	\includegraphics[width=0.8\columnwidth]{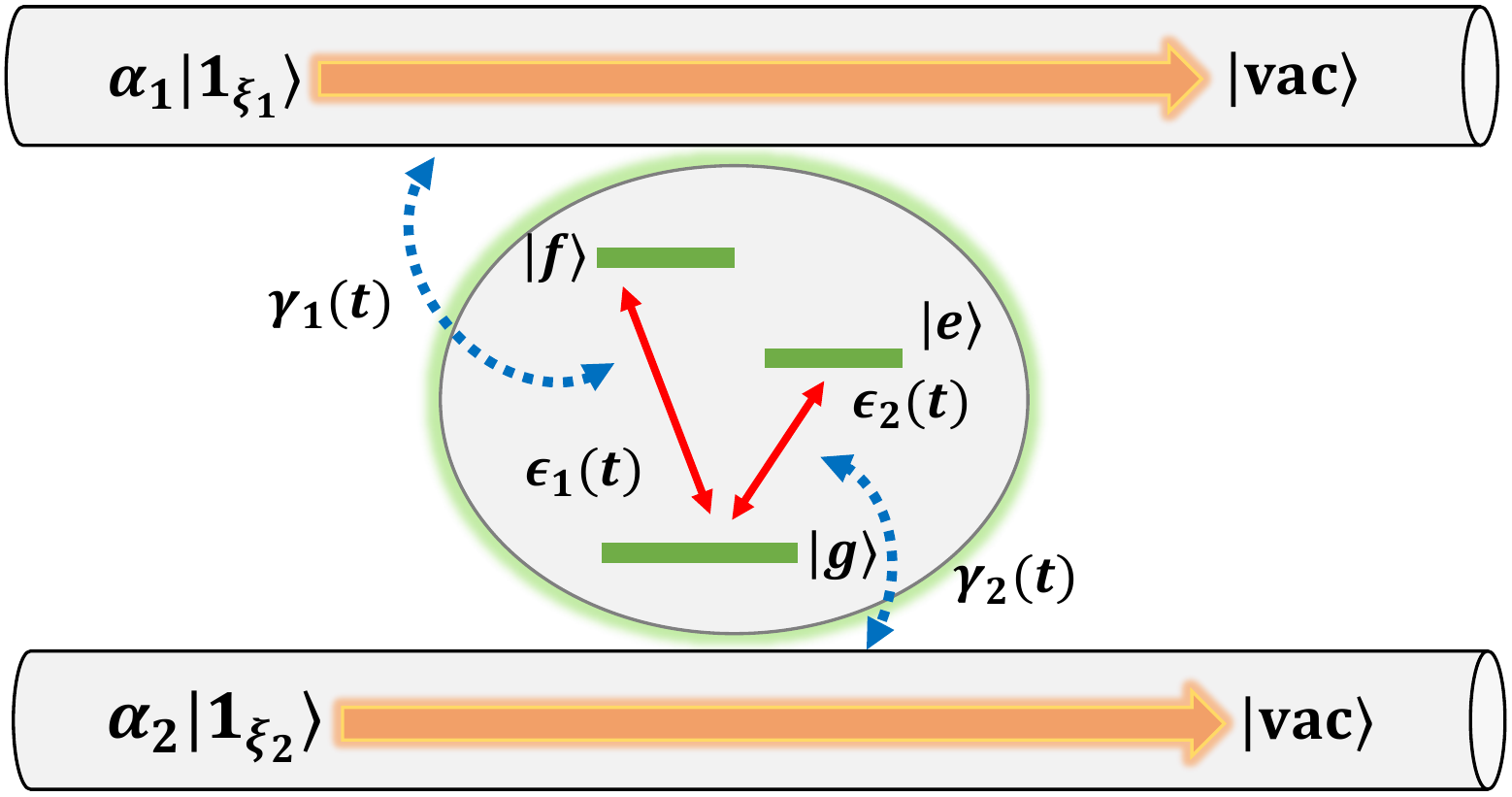}
 	\caption{Schematics for catching flying qubits at state $\alpha_1|1_{\xi_1}\rangle|{\rm vac}\rangle +\alpha_2|{\rm vac}\rangle|1_{\xi_2}\rangle$ by a $V$-type atom at state $|g\rangle$. Here, $\epsilon_1(t)$ and $\epsilon_2(t)$ are the detuning frequencies between the incident field and the $|f\rangle \leftrightarrow |g\rangle$ and $|f\rangle \leftrightarrow |e\rangle$ transistion frequencies, respectively. $\gamma_1(t)$ and $ \gamma_2(t)$ are the coupling functions between the atom and the channels.} \label{fig:8}
 \end{figure}

We can also use a $V$-type atom to catch a distributed single photon. As is shown in Fig.~\ref{fig:8}, the photon fields coming from both channels, forming an entangled state $\alpha_1|1_{\xi_1}\rangle|{\rm vac}\rangle +\alpha_2|{\rm vac}\rangle|1_{\xi_2}\rangle$, where $\xi_k(t)=|\xi_k(t)|e^{-i\phi_k(t)}$ ($k=1,2$). When the atom is initially prepared at the ground state $|g\rangle$, we can prove that the coupling functions and detuning frequencies for the entangled qubits to be caught by the three-level atom is
 \begin{equation}\label{eq:the conditions about gamma when catch by V}
 \gamma_{k}(t)=\frac{|\xi_k(t)|^2}{\int_{-\infty}^t|\xi_k(s)|^2 {\rm d}s},\quad \epsilon_k(t)=\dot{\phi}_k(t),
 \end{equation}	
where $k=1,2$ (the derivation is straightforward and will not be provided here). Consequently, the entangled state is transferred to the atom's superposition state $\alpha_1|f\rangle +\alpha_2|e\rangle$.

The obtained solutions can still be interpreted by the rule we observed in the above analysis, that is, the denominators are equal to the population of the associated excited state accumulated from the absorption of the corresponding single photon.

\section{The conversion of flying qubits by a three-level atom}\label{Sec:Transformation}
 	\begin{figure}
 		\centering
 		\includegraphics[width=0.8\columnwidth]{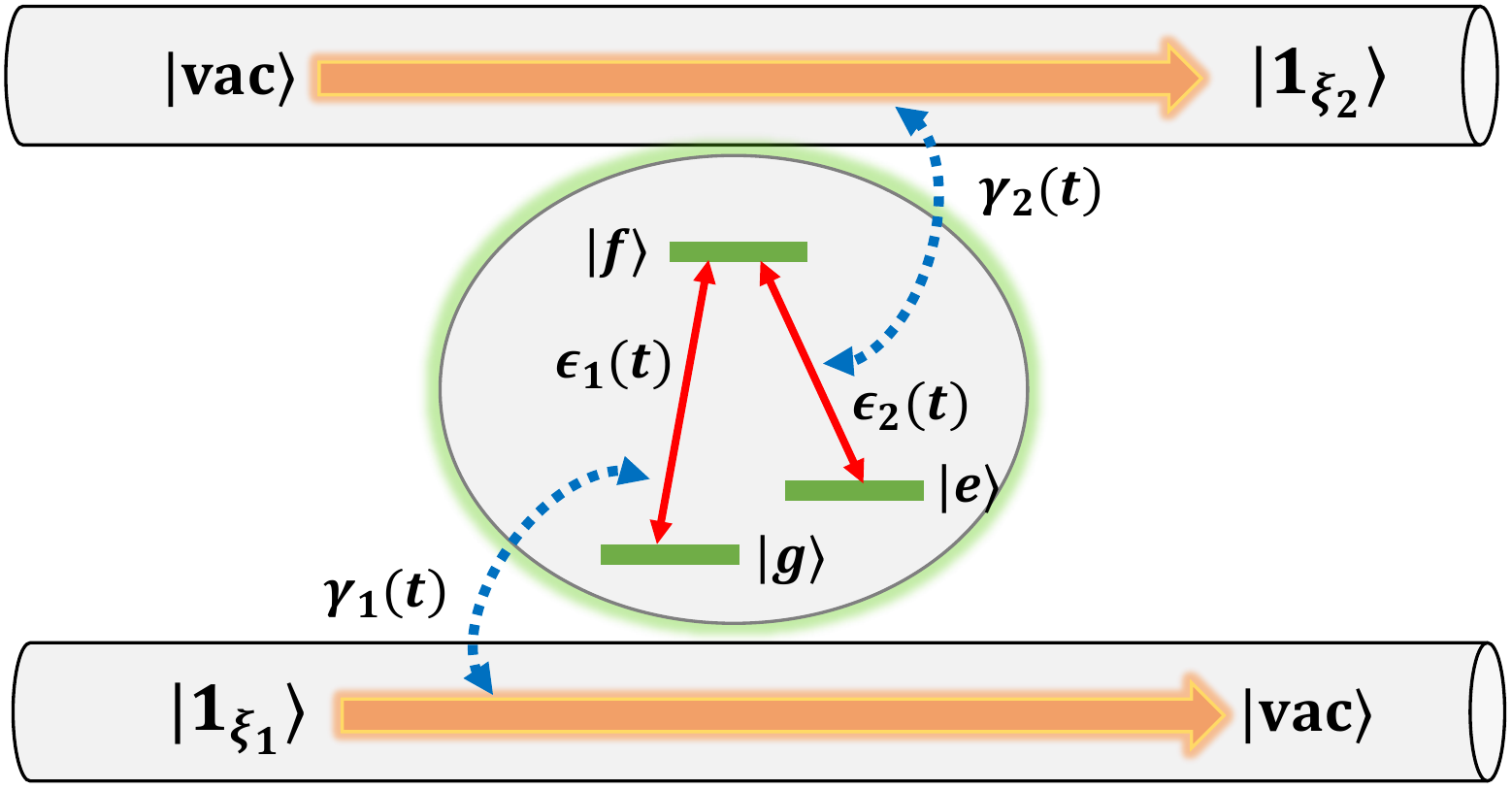}
 		\caption{Schematics for the conversion of flying qubits from state $|1_{\xi_1}\rangle|{\rm vac}\rangle$ to $|{\rm vac}\rangle|1_{\xi_2}\rangle$ by a $\Lambda$-type atom. Here, $\epsilon_1(t)$ and $\epsilon_2(t)$ are the detuning frequencies between the incident field and the $|f\rangle \leftrightarrow |g\rangle$ and $|f\rangle \leftrightarrow |e\rangle$ transition frequencies, respectively. $\gamma_1(t)$ and $ \gamma_2(t)$ are the coupling coefficients between the $\Lambda$-type atom and the channels.} \label{fig:9}
 	\end{figure}	

It is natural to use a $\Lambda$-type atom to convert a single photon between the two channels that it couples to. The conversion can not only reshape the single photon but also change its carrier frequency (note that $\xi(t)$ is the envelope of the single-photon shape function in the rotating-wave frame).

As is shown in Fig.~\ref{fig:9}, we want to design control protocols such that the atom converts a single-photon (at state $|1_{\xi_1}\rangle$) in the first channel to a single photon (at state $|1_{\xi_2}\rangle$) in the second channel. This consists of the catching of the first photon via transition between $|g\rangle$ and $|f\rangle$ and then the release of the second photon via transition between $|e\rangle$ and $|f\rangle$.

Similarly, we can introduce an ancillary system to generate $|1_{\xi_1}\rangle$ and obtain the conditions for the coupling functions and detuning frequencies to satisfy. In Appendix~\ref{conversion cascaded flying qubit}, we derive that
   \begin{equation} \label{eq:the conditions about gamma when switching by Lambda}
   \begin{split}
\gamma_1(t)&=\frac{|\xi_1(t)|^2}{\int_{-\infty}^t|\xi_1(\tau)|^2{\rm d}\tau-\int_{-\infty}^t|\xi_2(\tau)|^2{\rm d}\tau}, \\
\gamma_2(t)&=\frac{|\xi_2(t)|^2}{\int_t^\infty|\xi_2(\tau)|^2{\rm d}\tau-\int_t^\infty|\xi_1(\tau)|^2 {\rm d}\tau},
   \end{split}
 \end{equation}
and the detuning frequencies are
    \begin{equation} \label{eq:the conditions about epsilon when switching by Lambda}
    \epsilon(t)=\dot{\phi}_1(t)=\dot{\phi}_2(t).
    \end{equation}
Moreoever, for physically realizable conversions, the phase functions must satisfy
$$\phi_1(t)-\phi_2(t)=(2n+1)\pi,\quad  n\in Z.$$
This is different from the phase condition for the generation of single photons.

The obtained solution for flying-qubit conversion is also physical realizable under the condition (\ref{eq:constraint}) on the tail areas of $|\xi_1(t)|^2$ and $|\xi_2(t)|^2$. This is because the outgoing photon must be released after the incoming photon is caught, or equivalently saying, the radiated energy into the outgoing photon must be less than the energy absorbed from the incoming photon. Their difference, as described by the common denominator in Eq.~(\ref{eq:the conditions about gamma when switching by Lambda}), is nothing but the gained population of the state $|f\rangle$ shared by the two channels. This again conforms to the population rule we observed in the above three control scenarios.

Hence, the shape of the incoming photon $\xi_1(t)$ must decay more quickly than that of the outgoing photon, and consequently, $\gamma_1(t)$ will asymptotically approach to ${\bar{\bar \gamma}}_1(t)$ (the solution for catching $|1_{\xi_2}\rangle$ with a two-level atom) when $t\rightarrow -\infty$, and $\gamma_2(t)$ will approach to ${\bar{\bar \gamma}}_2(t)$ (the solution for generating $|1_{\xi_2}\rangle$ with a two-level atom) when $t\rightarrow \infty$.

\begin{figure}
	\centering
	\includegraphics[width=1\columnwidth]{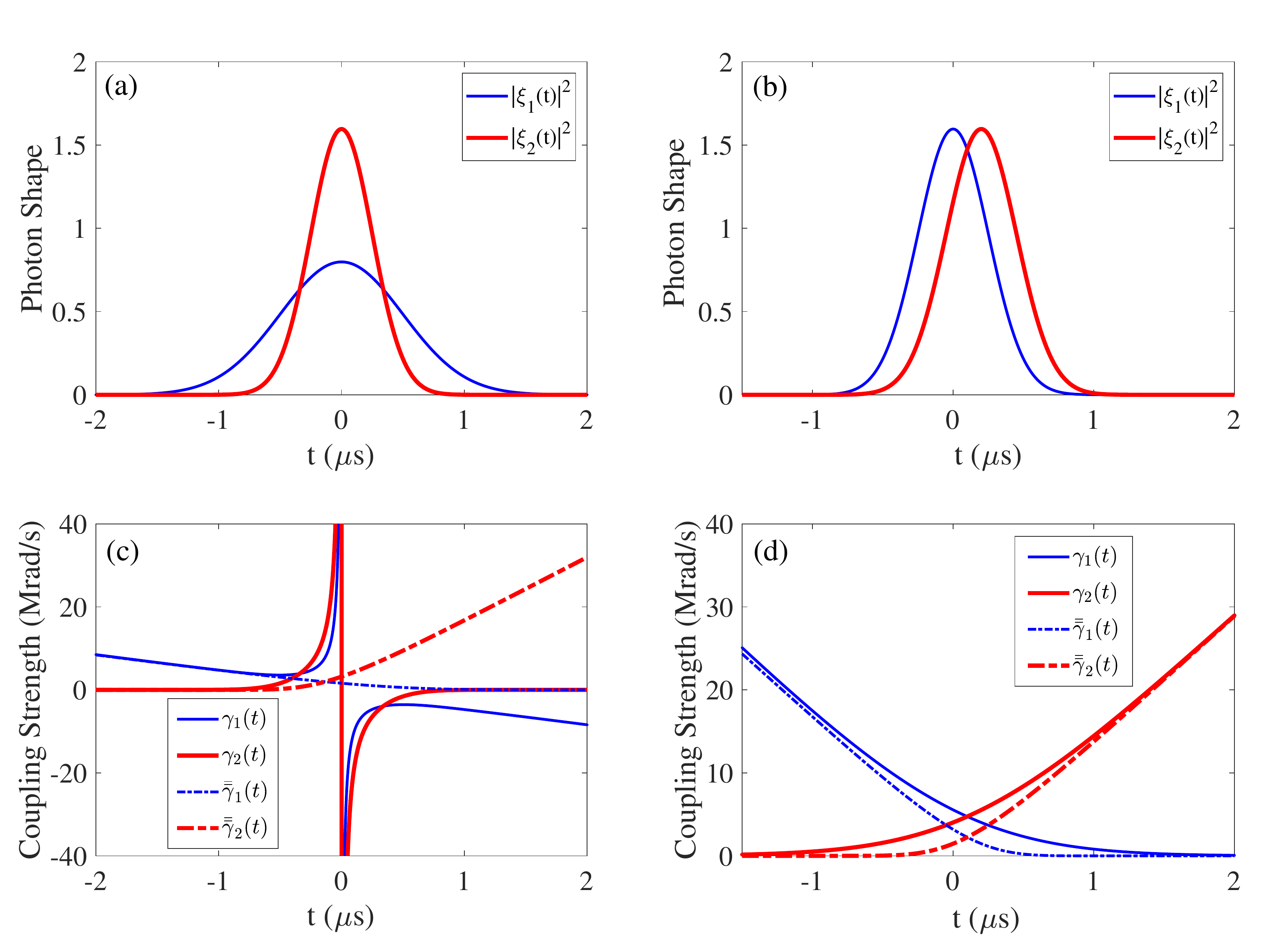}
	\caption{The conversion of a flying qubit $|1_{\xi_1}\rangle$ in the first channel into a flying qubit  $|1_{\xi_2}\rangle$ using a $\Lambda$-type atom. The incident photon shapes $\xi_1(t)$ and $\xi_2(t)$ are with (a) different Gaussian shapes or (b) same Gaussian shapes. The corresponding coupling strengths are shown in (c) and (d) respectively.} \label{fig:10}
\end{figure}
For verification, we test the same group of Gaussian shapes simulated in Sec.~\ref{Sec:Generation}. As is shown in Figs.~\ref{fig:10}(a) and (c), the violation of the tail area condition leads to non-physical negative coupling functions $\gamma_1(t)$ and $\gamma_2(t)$ after the peak time. In Figs~\ref{fig:10}(b) and (d), the resulting coupling functions are physically realizable, and one can clearly observe the predicted asymptotic limits of $\gamma_1(t)$ and $\gamma_2(t)$ when $t\rightarrow \pm \infty$.

 \section{Conclusion and outlook}\label{Sec:Conclusion}
To conclude, we have explored the control of flying qubits using various types of three-level atoms through tunable couplings to two input-output channels, including the generation of entangled qubits and correlated flying qubits, as well as the catching and conversion of single flying qubits. Analytic formulas are presented for the coupling and detuning functions to fulfil these control tasks, which lay a basis for systematic design of control protocols for high-fidelity flying-qubit transmission over complex quantum networks.

The method applied here can be naturally generalized to more complicated cases that include more atomic levels and input-output channels, as long as the number of excitation number (or energy) is preserved. We conjecture that, owing to the conservation law, the population rule concluded in this paper will still hold, i.e., the flying-qubit control problems can be decomposed into several single-photon generation or catching problems, and the corresponding coupling function is equal to the ratio between the square norm of the desired photon shape and the current population of the excited state associated with this photon. Whether this holds for most general cases is to be examined in our future studies.

In addition to the three types of three-level atoms we studied here, it is more intriguing to apply a $\Delta$-type atom, which is only realizable by aritificial atoms, for the control of flying qubits. For example, one can use it to convert a single photon into a pair of correlated photon pair, or vice versa, or routing a single photon between two different channels. However, the underlying scattering process is much more complicated, and we have not obtained any useful analytic solutions.

It should noted that the obtained solutions may confront physical constraints on the realizable shape functions $\xi_1(t)$ and $\xi_2(t)$, such as the phase conditions and the restriction (\ref{eq:constraint}) on the pulse tail areas. Towards more flexible flying-qubit shaping controls, one may have to introduce coherent controls that breaks the conservation of the number of excitations. Under such circumstances, analytic solutions are usually unavailable, but numerical optimization methods can be introduced to assist the design. All these problems will be explored in our future studies.

\section{Acknowledgments}\label{Sec:Acknowledgments}
This work is supported by the National Natural Science Foundation of China under Grant Nos. 61833010, 62173201 and 6217023269. GZ acknowledges supports from the Hong Kong Research Grant council (RGC) grants (Nos. 15203619 and 15208418), the Shenzhen Fundamental Research Program under the Grant
No. JCYJ20190813165207290, and the CAS AMSS-PolyU Joint Laboratory of Applied Mathematics.
 \appendix

\section{Proofs of main results}
For conciseness., we introduce several notations as follows:
\begin{eqnarray}
  \Gamma_j(t) &=& \int_{-\infty}^t \gamma_j(\tau){\rm d}\tau, \\
  \Theta_j(t) &=& \int_{-\infty}^t \epsilon_j(\tau){\rm d}\tau,
\end{eqnarray}
where $\gamma_j(\tau)$ and $\epsilon_j(\tau)$ ($j=1,2$) are the coupling and detuning functions of the $j$th channel. Based on the above notations, we further denote $\gamma(t)=\gamma_1(t)+\gamma_2(t)$ and $\epsilon(t)=\epsilon_1(t)+\epsilon_2(t)$.
$\Gamma(t)=\Gamma_1(t)+\Gamma_2(t)$ and $\Theta(t)=\Theta_1(t)+\Theta_2(t)$.

In the following, we will prove the main results presented in the main text.

\subsection{The derivation of Eq.~(\ref{eq:the conditions about gamma when generating by lambda}) and Eq.~(\ref{eq:the conditions about epsilon when generating by lambda})}\label{app:generationA}

Applying Eqs.~(\ref{eq:the effictive H})-(\ref{eq:the joint system-field state at infty}),  it is straightforward to find that, with the Hamiltonian and coupling operator given in Sec.~\ref{Sec:GenerationA}, only the single-photon emissions can occur when the three-level atom is initially prepared at $|f\rangle$ and ends up in $|g\rangle $ or $|e\rangle $. The calculation of Eq.~(\ref{eq:the final state}) on the single-photon emission gives
\begin{eqnarray}\label{}
\xi_1^g(\tau)&=&\sqrt{\gamma_1({\tau})}e^{-\frac{1}{2}\Gamma(s)-i\Theta(s)}, \\  \xi_2^e(\tau)&=&\sqrt{\gamma_2({\tau})}e^{-\frac{1}{2}\Gamma(s)-i\Theta(s)}.
\end{eqnarray}	

To generate the target distributed single photon that are entangled with the atom at state $\alpha_1|g\rangle|1_{\xi_1}\rangle|{\rm vac}\rangle + \alpha_2|e\rangle|{\rm vac}\rangle|1_{\xi_2}\rangle$, we should have $\xi_1^g(\tau)=\alpha_1\xi_1(\tau)$ and $\xi_2^e(\tau)=\alpha_2\xi_2(\tau)$, and hence
\begin{eqnarray}\label{}
\alpha_1\xi_1(\tau)&=&\sqrt{\gamma_1({\tau})}e^{-\frac{1}{2}\Gamma(s)-i\Theta(s)}, \label{eq:xi_1}\\  \alpha_2\xi_2(\tau)&=&\sqrt{\gamma_2({\tau})}e^{-\frac{1}{2}\Gamma(s)-i\Theta(s)}. \label{eq:xi_2}
\end{eqnarray}	
Let $\xi_k(\tau)=\left|\xi_k({\tau})\right|e^{-i\phi_k(t)}$, $k=1,2$. The phase condition Eq.~(\ref{eq:the conditions about epsilon when generating by lambda}) can be immediately obtained by comparing the phases of the left and right hand sides of Eqs.~(\ref{eq:xi_1})-(\ref{eq:xi_2}). These two equations also imply
\begin{equation}\label{eq:the ratio}
\frac{|\alpha_1\xi_1(\tau)|^2}{|\alpha_2\xi_2(\tau)|^2}=\frac{\gamma_1(\tau)}{\gamma_2(\tau)},
\end{equation}
and
\begin{equation}\label{eq:the sum}
\left|\alpha_1\xi_1({\tau})\right|^2+\left|\alpha_2\xi_2({\tau})\right|^2=\gamma(\tau)e^{-\Gamma(\tau)}=-\frac{\rm d}{{\rm d}\tau}e^{-\Gamma(\tau)}.
\end{equation}
The integration of both sides of Eq.~(\ref{eq:the sum}) from $-\infty$ to $t$ gives
\begin{equation}
\int_{-\infty}^t\left[\left|\alpha_1\xi_1({\tau})\right|^2{\rm d}\tau+\left|\alpha_2\xi_2({\tau})\right|^2\right]{\rm d}\tau =-e^{-\Gamma(\tau)},
\end{equation}
from which we solve
\begin{equation}\label{eq:gamma_B(t)}
\gamma(t)=\frac{\left|\alpha_1\xi_1(t)\right|^2+\left|\alpha_2\xi_2(t)\right|^2}{\int_{-\infty}^t\left[\left|\alpha_1\xi_1({\tau})\right|^2+\left|\alpha_2\xi_2({\tau})\right|^2 \right]{\rm d}\tau}.
\end{equation}
The solution (\ref{eq:the conditions about gamma when generating by lambda}) of $\gamma_1(t)$ and $\gamma_2(t)$ can thus be obtained by (\ref{eq:the ratio}) and (\ref{eq:gamma_B(t)}).

\subsection{The derivation of Eq.~(\ref{eq:the conditions about gamma when generating by Xi}) }\label{generator cascaded flying qubit}

We apply Eqs.~(\ref{eq:the effictive H})-(\ref{eq:the joint system-field state at infty}) with the model of $\Xi$-type atom described in Sec.~\ref{Sec:GenerationB}. It is found that, when the three-level atom is initially prepared at $|f\rangle$ and ends up in $|g\rangle $,  only the two-photon parts are left in Eq.~(\ref{eq:the final state}), as follows	
\begin{widetext}
	\begin{equation}\label{eq:}
	\xi_{1,2}^g(\tau_1,\tau_2)=\sqrt{\gamma_1({\tau_1})\gamma_2({\tau_2})}e^{-\frac{1}{2}\int_{-\infty}^{\tau_1}\gamma_1(s){\rm d}s-\frac{1}{2}\int_{\tau_1}^{\tau_2}\gamma_2(s){\rm d}s}e^{-i\int_{-\infty}^{\tau_1}\epsilon_1(s){\rm d}s-i\int_{\tau_1}^{\tau_2}\epsilon_2(s){\rm d}s}.
	\end{equation}	
\end{widetext}

According to the control target, the desired shape functions $|\xi_1({\tau_1})|^2$ and $|\xi_2({\tau_2})|^2$ should be the marginal distributions of $|\xi_{1,2}^g(\tau_1,\tau_2)|^2$, respectively, from which we derive that
\begin{equation}\label{eq:the xi_1}
|\xi_1({\tau_1})|^2=\int_{\tau_1}^{\infty}|\xi_{1,2}^g(\tau_1,\tau_2)|^2{\rm d}\tau_2=\gamma_1({\tau_1})e^{-\Gamma_1(\tau_1)},
\end{equation}
and
\begin{equation}\label{eq:the xi_2}
\begin{split}
|\xi_2({\tau_2})|^2&=\int_{-\infty}^{\tau_2}|\xi_{1,2}^g(\tau_1,\tau_2)|^2{\rm d}\tau_1
\\&=\gamma_2(\tau_2)e^{-\Gamma_2(\tau_2)}\int_{-\infty}^{\tau_2}|\xi_1({\tau_1})|^2e^{\Gamma_2(\tau_1)}{\rm d}\tau_1.
\end{split}
\end{equation}

Equation~(\ref{eq:the xi_1}) can be simplified to obtain Eq.~(\ref{eq:the conditions about gamma when generating by Xi}a) for the solution of $\gamma_1(t)$. As for the solution of $\gamma_2(t)$, we can rewrite Eq.~(\ref{eq:the xi_2}) as
\begin{equation}\label{eq:the relation between gammB1 and xi1, xi2}
\int_{-\infty}^{\tau_2}|\xi_1({\tau_1})|^2e^{\Gamma_2(\tau_1)}{\rm d}\tau_1=\frac{|\xi_2({\tau_2})|^2}{\gamma_2(\tau_2)} e^{\Gamma_2(\tau_2)},
\end{equation}
and differentiate both sides of Eq.~(\ref{eq:the relation between gammB1 and xi1, xi2}) to obtain the solution (\ref{eq:the conditions about gamma when generating by Xi}b).

\subsection{The derivation of Eq.~(\ref{eq:the conditions about gamma when switching by Lambda}) and Eq.~(\ref{eq:the conditions about epsilon when switching by Lambda})}\label{conversion cascaded flying qubit}
We first construct an auxiliary system $A$ to virtually generate the incident single photon $|1_{\xi_1}\rangle$, for which one can set
\begin{equation}\label{}
  H_A(t) = 0,\quad L_A(t)=\sqrt{\gamma_A(t)}\sigma_-,
\end{equation}
with the coupling function being
\begin{equation}\label{}
\gamma_A(t)=\frac{|\xi_1(t)|^2}{\int_t^\infty|\xi_1(s)|^2{\rm d}s},
\end{equation}
where, $\gamma_A(t)$ is the coupling function between the auxiliary system and first channel.
According to the $(S,L,H)$ formula~\cite{Gough2009}, the equivalent Hamiltonian of the joint system as Eq.~(\ref{eq:the equivalent Hamiltonian of the joint system}) is
\begin{equation}\label{eq:equivalent Hamiltonian of capture}
 \begin{split}
\bar{H}(t)&= \mathbb{I}_A\otimes\left[\epsilon_1(t)+\epsilon_2(t)\right]|f\rangle \langle f|
\\&+\frac{1}{2i}\sqrt{\gamma_A(t)\gamma_1(t)}\left[|g,f\rangle \langle e,g|  -H.c\right],
 \end{split}
\end{equation}
where the notation $|\alpha,\beta\rangle=|\alpha\rangle\otimes|\beta\rangle$ is adopted. The corresponding equivalent coupling operators as Eq.~(\ref{eq:the equivalent coupling operators of the joint system}) are
\begin{equation}\label{eq:equivalent coupling operator L_1 of capture}
 \begin{split}
\bar{L}_1(t)&=\mathbb{I}_A\otimes\sqrt{\gamma_1(t)}|g\rangle \langle f| +\sqrt{\gamma_A(t)}|g\rangle \langle e|\otimes\mathbb{I} ,\\
\bar{L}_2(t)&=\mathbb{I}_A\otimes\sqrt{\gamma_2(t)}|e\rangle \langle f|.
 \end{split}
\end{equation}

 Applying Eq.~(\ref{eq:the solution of the equivalent state of system}) with the above equivalent $\bar{H}(t)$ and $\bar{L}_j(t)$, we can see that only the following three components of correlated system states are non-vanishing
 \begin{equation}\label{eq:the every state in solution}
 \begin{split}
 &|\psi(t)\rangle=e^{-\frac{1}{2}\Gamma_A(t)}|e,g\rangle -\Xi(t)e^{-\frac{1}{2}\Gamma(t)}|g,f\rangle,
 \\&|\psi_1(\tau|t)\rangle=\left[\xi_1(\tau)-\sqrt{\gamma_{1}({\tau})}e^{-\frac{1}{2}\Gamma(\tau)-i\Theta(\tau)}\Xi({\tau})\right]|g,g\rangle,
 \\&|\psi_2(\tau|t)\rangle=-\sqrt{\gamma_2({\tau})}e^{-\frac{1}{2}\Gamma(\tau)-i\Theta(\tau)}\Xi({\tau})|g,e\rangle,
 \end{split}
 \end{equation}
where $\Gamma_A(t)=\int_{-\infty}^{t} \gamma_A(s){\rm d}s$ and
 \begin{equation}\label{eq:the condition about Xi and Gamma}
 \Xi(t)=\int_{-\infty}^{t}\sqrt{\gamma_1(\tau)}e^{\frac{1}{2}\Gamma(\tau)+i\Theta(\tau)}\xi_1(\tau){\rm d}\tau.
 \end{equation}
The three non-vanishing components correspond to the cases when both channels have vacuum output, the first channel has a single-photon output, and the second channel has a single-photon output, respectively.

Note that $|\psi_1(\tau|t)\rangle$ and $|\psi_2(\tau|t)\rangle$ are both independent of $t$, and hence we have
\begin{eqnarray}
\xi_1^g(\tau)\!&=&\!\xi_1(\tau)-\sqrt{\gamma_{1}({\tau})}e^{-\frac{1}{2}\Gamma(\tau)-i\Theta(\tau)}\Xi({\tau}),\label{eq:the solution about x_1 about switch} \\
\xi_2^e(\tau)\!&=&\!-\sqrt{\gamma_2({\tau})}e^{-\frac{1}{2}\Gamma(\tau)-i\Theta(\tau)}\Xi({\tau}),\label{eq:the solution about x_2 about switch}
\end{eqnarray}
To convert the incident flying qubit $|1_{\xi_1}\rangle$ into the outgoing flying qubit $|1_{\xi_2}\rangle$, the single photon must completely go out through the second channel, and hence $\xi_1^g(\tau)=0$ and $\xi_1^e(\tau)=\xi_2(\tau)$, which implies that the two shape functions satisfy
\begin{eqnarray}
\xi_1(\tau)&=&\sqrt{\gamma_{1}({\tau})}e^{-\frac{1}{2}\Gamma(\tau)-i\Theta(\tau)}\Xi({\tau}), \label{eq:the conditions about xi_1 when switching by Lambda}\\
\xi_2(\tau)&=&-\sqrt{\gamma_2({\tau})}e^{-\frac{1}{2}\Gamma(\tau)-i\Theta(\tau)}\Xi({\tau}).\label{eq:the conditions about xi_2 when switching by Lambda}
\end{eqnarray}

Now we apply the normalization condition of the atom-field joint state (\ref{eq:the joint system-field state}), which requires that
 \begin{equation}\label{eq:the law of conservation of the number of excited states I}
 \begin{split}
  1  & \equiv    \langle \Psi(t)|\Psi(t)\rangle  =    \langle \psi(t)|\psi(t)\rangle \\
   & \quad + \int_{-\infty}^t
 \left[ \langle\psi_1(\tau|t)|\psi_1(\tau|t)\rangle +\langle \psi_2(\tau|t)|\psi_2(\tau|t)\rangle \right] {\rm d}\tau.
 \end{split}
 \end{equation}
Combining with Eqs.~(\ref{eq:the every state in solution}), (\ref{eq:the conditions about xi_1 when switching by Lambda}) and (\ref{eq:the conditions about xi_2 when switching by Lambda}), we can transform Eq.~(\ref{eq:the law of conservation of the number of excited states I}) as follows:
\begin{equation}\label{eq:the law of conservation of the number of excited states III}
  1 \equiv \int_t^{\infty} |\xi_1(\tau)|^2{\rm d}\tau + |\Xi(t)|^2 e^{-\Gamma(t)} + \int_{-\infty}^t  |\xi_2 (\tau)|^2 {\rm d}\tau,
  \end{equation}
where the relation $e^{-\Gamma_A(t)}=\int_t^{\infty} |\xi_1(s)|^2{\rm d}s$ is used.

This equation precisely describes the conservation of excitation number (or energy in the system). The system is initially prepared at the ground state with a single-photon input, and hence its total number of excitations is 1, as indicated by the left hand side  of Eq.~(\ref{eq:the law of conservation of the number of excited states III}). At any time $t>0$, the energy contained in the photon may stay in the input field (the first term on the right hand side) or be transferred to the population of $|f\rangle$ (the second term on the right hand side ) and the output field (the third term on the right hand side). When $t\rightarrow \infty$, all the energy flows form the incoming photon to the outgoing photon, while the first and second terms decay to zero.

Further, Eq.~(\ref{eq:the law of conservation of the number of excited states III}) implies that
\begin{equation}\label{eq:|Xi|2}
\begin{split}
|\Xi(t)|^2 e^{-\Gamma(t)}& = \int^t_{-\infty} |\xi_1(\tau)|^2{\rm d}\tau  -\int_{-\infty}^t  |\xi_2 (\tau)|^2 {\rm d}\tau,
\\& =  \int_t^{\infty} |\xi_2(\tau)|^2{\rm d}\tau  -\int^{\infty}_t  |\xi_1 (\tau)|^2 {\rm d}\tau,
\end{split}
\end{equation}
as both $\xi_1(\tau)$ and $\xi_2(\tau)$ are normalized functions. Replacing this equation back into Eqs.~(\ref{eq:the conditions about xi_1 when switching by Lambda}) and (\ref{eq:the conditions about xi_2 when switching by Lambda}), we obtain the solutions for the coupling functions:
   \begin{equation} \label{eq:}
   \begin{split}
\gamma_1(t)&=\frac{|\xi_1(t)|^2}{\int_{-\infty}^t|\xi_1(\tau)|^2{\rm d}\tau-\int_{-\infty}^t|\xi_2(\tau)|^2{\rm d}\tau}, \\
\gamma_2(t)&=\frac{|\xi_2(t)|^2}{\int_t^\infty|\xi_2(\tau)|^2{\rm d}\tau-\int_t^\infty|\xi_1(\tau)|^2 {\rm d}\tau}.
   \end{split}
 \end{equation}

As for the solutions of the detuning functions, one can directly observe that Eqs.~(\ref{eq:the conditions about xi_1 when switching by Lambda}) and (\ref{eq:the conditions about xi_2 when switching by Lambda}) are satisfied when
\begin{equation}\label{eq:the conditions about epsilon when switching by Lambda I}
\epsilon(t)=\dot{\phi}_1(t), \quad \epsilon(t)=\dot{\phi}_2(t),
\end{equation}
where $\phi_1(t)$ and $\phi_2(t)$ are the phase functions of $\xi_1(t)$ and $\xi_2(t)$, respectively. Moreover, Eqs.~(\ref{eq:the conditions about xi_1 when switching by Lambda}) and (\ref{eq:the conditions about xi_2 when switching by Lambda}) indicates that $\xi_1(t)$ and $\xi_2(t)$ must have inverted phases, i.e.,
$$\phi_1(t)-\phi_2(t)=(2n+1)\pi,\quad  n\in Z.$$

The obtained equations also provide the solution for catching an incident single photon from the first channel, which requires that $\xi_1^g(\tau)=\xi_2^e(\tau)=0$. According to Eq.~(\ref{eq:the conditions about xi_2 when switching by Lambda}), $\gamma_2(t)$ must be turned off so that the resulting design is equivalent to that with a two-level atom.


\end{document}